\documentstyle[prl,preprint,aps,epsf]{revtex}

\begin{document}
\bibliographystyle{plain}
\title{Condition for Superradiance in Higher-dimensional Rotating
Black Holes}
\author{Eylee Jung\footnote{Email:eylee@kyungnam.ac.kr}, 
SungHoon Kim\footnote{Email:shoon@kyungnam.ac.kr} and
D. K. Park\footnote{Email:dkpark@hep.kyungnam.ac.kr 
}}
\address{Department of Physics, Kyungnam University,
Masan, 631-701, Korea.}
\date{\today}
\maketitle

\begin{abstract}
It is shown that the superradiance modes always exist in the radiation by the 
$(4+n)$-dimensional rotating black holes. Using a Bekenstein argument the 
condition for the superradiance modes is shown to be 
$0 < \omega < m \Omega$ for the scalar, electromagnetic and gravitational 
waves when the spacetime background has a single angular momentum 
parameter about an axis on the brane, where $\Omega$ is a rotational frequency
of the black hole and $m$ is an azimuthal quantum number of the radiated
wave.
\end{abstract}
\newpage
Recent brane-world scenarios such as large extra 
dimensions\cite{ark98-1,anto98} or compactified extra dimensions with warped
factor\cite{rs99-1} predict a TeV-scale gravity. The emergence for the 
TeV-scale gravity in the higher-dimensional theories opens the possibility
to make the black hole factories in the future high-energy
colliders\cite{gidd02-1,dimo01-1,eard02-1,stoj04}. In this context it is
of interest to examine the various properties of the higher-dimensional 
black holes.

The absorption and emission for the different particles by the 
($4+n$)-dimensional Schwarzschild background have been studied 
analytically\cite{jung04} and numerically\cite{harris03-1}. It was shown that
the presence of the extra dimensions in general decreases the absorptivity
and increases the emission rate on the brane. The decrease of the absorptivity
may be due to the decrease of the effective radius\cite{emp00} 
$r_c \equiv \sqrt{\sigma_{\infty} / \pi}$, where $\sigma_{\infty}$ is a 
high-energy limit of the total absorption cross section. Although it may explain
why the absorptivity is suppressed in the high-energy regime, it does not
seem to provide a satisfactory physical reason for the suppression of the 
absorptivity in the full range of the particle energy. The enhancement of the
emission rate may be caused by the increase of the Hawking temperature in
the presence of the extra dimensions. This means that the Planck factor is more
crucial than the greybody factor in the Hawking radiation. For the case
of the minimally coupled massless scalar the low-energy absorption cross
section(LACS) always equals to the horizon area\cite{das97}. 
Thus, for the brane-localized scalar the LACS is always $4 \pi r_H^2$ while
for the bulk scalar it is equals to $\Omega_{n+2} r_H^{n+2}$ where 
$\Omega_{n+2} = 2 \pi^{(n+3) / 2} / \Gamma[(n+3) / 2]$ is the area of a unit
$(n+2)$-sphere. The ratio of the LACS for the Dirac field to that for the 
scalar was shown to be $2^{(n-3)/(n+1)}$ for the brane-localized 
case\cite{jung04} and $2^{-(n+3) / (n+1)}$ for the bulk case\cite{jung05-1}. 
Therefore, the ratio factor $1/8$, which was obtained by Unruh long
ago\cite{unruh76}, was recovered when $n=0$. The dependence on the 
dimensionality in these ratio factors may be used to prove the existence of 
the extra dimensions in the future black hole experiments. 
The relative bulk-to-brane
energy emissivity was also calculated in Ref.\cite{harris03-1} numerically, 
which confirmed the main result of Ref.\cite{emp00}, {\it i.e. black holes
radiate mainly on the brane}.

For the higher-dimensional charged black holes the full absorption and emission
spectra have been computed numerically in Ref.\cite{jung05-2}. It has been 
shown that contrary to the effect of the extra dimension the presence of
the nonzero inner horizon parameter $r_-$ generally enhances the absorptivity
and suppresses the emission rate. It has been shown also that the relative 
bulk-to-brane emissivity decreases with increasing the inner horizon parameter
$r_-$. The LACS for the minimally coupled massless scalar always equals to
the horizon area. For the Dirac fermion the LACS becomes \cite{jung05-1}
\begin{equation}
\label{fermionbl}
\sigma_F^{BL} = 2^{- \frac{n+3}{n+1}} \left[ 1 - \left( \frac{r_-}{r_+}\right)
^{n+1} \right]^{\frac{n+2}{n+1}} \sigma_S ^{BL}
\end{equation}
for the bulk case and
\begin{equation}
\label{fermionbr}
\sigma_F^{BR} = 2^{\frac{n-3}{n+1}} \left[ 1 - \left( \frac{r_-}{r_+}\right)
^{n+1} \right]^{\frac{2}{n+1}} \sigma_S ^{BR}
\end{equation}
for the brane-localized case. In Eqs. (\ref{fermionbl}) and (\ref{fermionbr}) 
$\sigma_S^{BL}$ and $\sigma_S^{BR}$ are the LACSs for the bulk and 
brane-localized scalars, respectively.

The absorption and emission problems in the higher-dimensional rotating black 
holes were recently discussed in Ref. \cite{frol03-1,ida05,harris05-1}, where
the existence of the superradiance modes \cite{zeldo71-1,press72} is
predicted analytically and numerically in the presence of the extra dimensions.
The existence of the superradiance is very important for the experimental
signature in the future colliders because it may 
change\cite{frol03-1,frol02-1,frol02-2}
the standard claim
that {\it black holes radiate mainly on the brane}. In this context it is 
important to derive a criterion for the existence of the superradiance. 
In this short note we will derive this criterion  using a 
Bekenstein's argument\cite{beken73}. 
 
The gravitational background around a $(4+n)$-dimensional, rotating, uncharged
black hole having single angular momentum parameter about an axis in the
brane is given by\cite{myers86}
\begin{eqnarray}
\label{metric1}
ds^2 = &-& \left( 1 - \frac{\mu}{\Sigma r^{n-1}} \right) dt^2 - \frac{2a\mu\sin^2\theta}{\Sigma r^{n-1}}
\,dt d\phi + \frac{\Sigma}{\bigtriangleup} dr^2 + \Sigma \,d\theta^2 \\
&+& \left( r^2 + a^2 + \frac{a^2\mu\sin^2\theta}{\Sigma r^{n-1}} \right) \sin^2\theta \,d\phi^2 + r^2
\cos^2 \theta \,d\Omega_n \nonumber
\end{eqnarray}
where
\begin{equation}
\label{sik1}
\bigtriangleup = r^2 + a^2 - \frac{\mu}{r^{n-1}},\hspace{1cm} \Sigma = r^2 + a^2 \cos^2 \theta ,
\end{equation}
and $d\Omega_n$ is a line-element on a unit $n$-sphere.

It is worthwhile noting that the $(4+n)$-dimensional rotating black holes can
have $1 + n/2$ angular momentum parameters for 
even $n$ and $(3+n) / 2$ parameters for odd $n$ maximally\cite{myers86}. 
Although our following argument can be applicable to this general case,
it seems to be complicated in the calculation. Thus, we would like to consider 
the simpler case by reducing the angular momentum parameters. That is why
we choose a single angular momentum parameter in Eq. (\ref{metric1}).  
The detailed calculation for the spacetime 
bakground having multiple angular momentum parameters will be reported
elsewhere.

The horizon radius $r_H$ is determined from $\bigtriangleup = 0$,\,{\it i.e.}
\begin{equation}
\label{sik2}
r_H ^2 + a^2 - \frac{\mu}{r_H ^{n-1}} = 0.
\end{equation}
The horizon area $\tilde{A}$, mass $M$, angular momentum $J$ and Hawking 
temperature $T_H$ are given by
\begin{eqnarray}
\label{sik3}
\tilde{A}&=&\frac{\Omega_n r_H ^n}{n+1} A \hspace{1.4cm}
M=\frac{(n+2)\Omega_{n+2}}{16\pi}\mu \\
J&=&\frac{2}{n+2}Ma    \hspace{1cm}
T_H = \frac{2}{A} \left[ r_H + \frac{8\pi(n-1)M}{(n+2)\Omega_{n+2}r_H ^n}  
                                                        \right]
\nonumber
\end{eqnarray}
where $\Omega_N = 2\pi^{(N+1)/2}/\Gamma[(N+1)/2]$ is an area of unit 
$N$-sphere and 
$A = 4\pi (r_H ^2 + a^2 )$. 
It is easy to show
that the various quantities in Eq. (\ref{sik3}) are related to each other 
in the form
\begin{equation}
\label{sik4}
AT_H = 2 r_H + (n-1)\frac{\mu}{r_H ^n} = 2r_H + \frac{(n-1)A}{4\pi r_H}.
\end{equation}
Now we assume $M$ and $J$ are independent variables. Then elementary 
mathematics gives
\begin{equation}
\label{sik5}
d\tilde{A} = \frac{\partial\tilde{A}}{\partial M}dM + \frac{\partial\tilde{A}}{\partial J}dJ.
\end{equation}
Firstly, let us calculate $\partial A /\partial M$, which is given by
\begin{equation}
\label{sik6}
\frac{\partial A}{\partial M} = 8\pi \left( r_H \frac{\partial r_H }{\partial M}
                                   + a \frac{\partial a}{\partial M} \right).
\end{equation}
Differentiating Eq. (\ref{sik2}) with respect to $M$ and using Eq. (\ref{sik4}), 
it is easy to show
\begin{eqnarray}
\label{sik7}
\frac{\partial r_H }{\partial M} &=& \frac{1}{AT_H} \left[ \frac{16\pi}{(n+2) 
     \Omega_{n+2}r_H ^{n-1}}
                      + \frac{2a^2}{M} \right] \\
\frac{\partial a }{\partial M} &=& - \frac{n+2}{2M^2}J = -\frac{a}{M}, \nonumber
\end{eqnarray}
which results in 
\begin{equation}
\label{sik8}
\frac{\partial A }{\partial M} = \frac{2}{M T_H r_H } \left[ (r_H ^2 + a^2 ) 
- na^2\right].
\end{equation}
Combining Eq. (\ref{sik7}) and (\ref{sik8}), one can show
\begin{equation}
\label{sik9}
\frac{\partial\tilde{A} }{\partial M} = 
\frac{(n+2)\Omega_n r_H ^{n-1}}{(n+1)MT_H}(r_H ^2 +a^2 ).
\end{equation}
Differentiating Eq. (\ref{sik2}) with respect to $J$ and following the previous 
procedure, one also
can show
\begin{eqnarray}
\label{sik10}
\frac{\partial A}{\partial J} &=& \frac{(n-1)(n+2)a}{MT_H r_H} \\
\frac{\partial\tilde{A}}{\partial J} &=& 
- \frac{(n+2)\Omega_n r_H ^{n-1}a}{(n+1)MT_H}. \nonumber
\end{eqnarray}
Inserting Eq. (\ref{sik9}) and (\ref{sik10}) into (\ref{sik5}), 
Eq. (\ref{sik5}) becomes in the following
\begin{equation}
\label{sik11}
d\tilde{A} = \left[ 1 - \Omega \frac{dJ}{dM} \right]
\frac{\partial\tilde{A}}{\partial M}dM
\end{equation}
where
\begin{equation}
\label{sik12}
\Omega = \frac{a}{r_H ^2 + a^2}
\end{equation}
is a rotational frequency of the black hole.
Bekenstein showed in Ref.\cite{beken73} that for scalar, electromagnetic and 
gravitational waves $dJ / dM$ becomes
\begin{equation}
\label{sik13}
\frac{dJ}{dM} = - \frac{T^r _\phi}{T^r _t} = \frac{m}{\omega}
\end{equation}
where $m$ and $\omega$ are azimuthal quantum number and energy of the 
incident wave respectively, and
$T_{\mu\nu}$ is a stress-energy tensor.
Thus Eq. (\ref{sik11}) becomes
\begin{equation}
\label{sik14}
d\tilde{A} = \left[ 1 - \frac{m}{\omega}\Omega  \right] \frac{\partial\tilde{A}}{\partial M}dM
\end{equation}
Since $\partial\tilde{A} /\partial M$ is always 
positive from Eq. (\ref{sik9}) and $d\tilde{A} > 0$ because
$\tilde{A} /4$ is a black hole entropy, Eq. (\ref{sik14}) gives a condition
\begin{equation}
\label{sik15}
0 < \omega< m\Omega
\end{equation}
if $dM<0$, which is a condition for the existence of the superradiance. 

One may apply the same procedure 
to the brane-localized fields to derive a criterion for the existence of 
the superradiance modes.
In this case we should use the induced metric
\begin{eqnarray}
\label{sik16}
ds^2 _{BR}= &-& \left( 1 - \frac{\mu}{\Sigma r^{n-1}} \right) dt^2 - \frac{2a\mu\sin^2\theta}{\Sigma r^{n-1}}
\,dt d\phi + \frac{\Sigma}{\bigtriangleup} dr^2 \\
&+& \Sigma \,d\theta^2 + \left( r^2 + a^2 + \frac{a^2\mu\sin^2\theta}{\Sigma r^{n-1}} \right) 
\sin^2\theta \,d\phi^2 . \nonumber
\end{eqnarray}
However, the metric (\ref{sik16}) is not exact black hole solution of the 
Einstein field equation. Thus
it is obscure whether we can identify the quarter of the horizon area 
with a black hole entropy.
Since, furthermore, the metric (\ref{sik16}) is not a vacuum solution unlike
Kerr black hole, it generates its own stress-energy tensor 
and hence total energy-momentum
tensor should be $T_{\mu\nu} ^{tot} = T_{\mu\nu} ^f +T_{\mu\nu} ^{m}$ 
where $T_{\mu\nu} ^f$ and 
$T_{\mu\nu} ^{m}$ are the stress-energy tensors contributed from field 
and metric, respectively. Thus,
it is not evident for the brane-localized waves whether we can use 
$-T_{\phi} ^{r,tot} /T_{t} ^{r,tot} = m/\omega$ or not. 

As commented earlier, our procedure is not restricted to the rotating 
black hole with single angular
parameter. In $(4+n)$-dimension the rotating black holes can have $(n+3)/2$ 
angular momentum parameters for odd $n$ and $(n+2)/2$ parameters for even $n$.
So it is interesting to apply our method to this  
black holes to derive a general condition for the
existence of the superradiance.

\vspace{1cm}

{\bf Acknowledgement}:  
This work was supported by the Kyungnam University
Research Fund, 2004.

\end{document}